\documentclass[reprint,superscriptaddress,amsmath,amssymb,prb,floatfix,]{revtex4-1}
\usepackage{graphicx}
\usepackage{dcolumn}
\usepackage{bm}
\usepackage{xcolor,soul}
\soulregister\cite7 
\soulregister\citep7 
\soulregister\citet7 
\soulregister\ref7 
\soulregister\pageref7 
\sethlcolor{green} 

\begin{document}


\title{Comment on ``Piezoelectricity in planar boron nitride via a geometric phase''}

\author{Jie Li}
\affiliation{School of Engineering, Sun Yat-sen University, Guangzhou 510006, China}

\author{Yunhua Wang}
\email{wangyh49@mail.sysu.edu.cn}
\affiliation{Sino-French Institute of Nuclear Engineering and Technology, Sun Yat-sen University, Zhuhai 519082, China}

\author{Zongtan Wang}
\affiliation{School of Engineering, Sun Yat-sen University, Guangzhou 510006, China}

\author{Jie Tan}
\affiliation{Sino-French Institute of Nuclear Engineering and Technology, Sun Yat-sen University, Zhuhai 519082, China}

\author{Biao Wang}
\email{wangbiao@mail.sysu.edu.cn}
\affiliation{Sino-French Institute of Nuclear Engineering and Technology, Sun Yat-sen University, Zhuhai 519082, China}
\affiliation{State Key Laboratory of Optoelectronic Materials and Technologies, School of Physics, Sun Yat-sen University, Guangzhou 510275, China}

\author{Yulan Liu}
\email{stslyl@mail.sysu.edu.cn}
\affiliation{School of Engineering, Sun Yat-sen University, Guangzhou 510006, China}


\begin{abstract}
Using the strain-dependent effective Hamiltonian and the geometric phase, Droth  \emph{et al}. [Phys. Rev. B \textbf{94}, 075404 (2016)] obtain an analytical expression for the electronic contribution to the piezoelectricity of planar hexagonal boron nitride (\emph{h}-BN). Their analytical results of piezoelectric constants for \emph{h}-BN are invalid because of the mistakes in constructing the adiabatic process of the piezoelectricity. In this comment, we reconstruct a proper adiabatic process for piezoelectricity and formulate a general Berry phase expression for the piezoelectric coefficients of two-dimensional piezoelectric crystals by means of the modern theory of polarization. The corrected analytical results of the piezoelectric constants are in complete consistency with the first-principles calculations and hence manifest the validity and generality of the Berry phase expression of piezoelectric coefficients.
\end{abstract}

\pacs{Valid PACS appear here}
\maketitle

Recently, Droth \emph{et al}. analytically evaluate the electronic contribution to the piezoelectricity of planar boron nitride (\emph{h}-BN) by means of the geometric phase of wave functions and the strain-dependent effective Hamiltonian of \emph{h}-BN \cite{droth2016piezoelectricity}. The analytical value of piezoelectric constant $e_{222}$ is $1.0\times10^{-10}C/m$, as shown in Eq. (14) of Ref.~[\onlinecite{droth2016piezoelectricity}]. However, the value of $e_{222}$ obtained from first-principles calculations is $3.71\times10^{-10}C/m$ in the case of clamped-ion models \cite{duerloo2012intrinsic}. Such large deviation of their analytical values from the first-principles calculations arises from a serious mistake in constructing the adiabatic process of the piezoelectricity. Consequently, the erroneous usage of the polarization difference formula in Eq. (4) leads to an invalid analytical expression of piezoelectric constant in Eq. (12) of Ref.~[\onlinecite{droth2016piezoelectricity}]. In this comment, we reconstruct a proper adiabatic process for piezoelectricity and present a general expression of the piezoelectric coefficients for two-dimensional (2D) piezoelectric crystals. For \emph{h}-BN, the obtained analytical result agrees well with the first-principles calculations.

The modern theory of polarization offers further insight on the physical understanding of dielectricity, piezoelectricity and ferroelectricity of crystalline systems \cite{king1993theory,resta1994macroscopic,PhysicsofFerroelectrics}, by virtue of the clear definition of the polarization difference, \emph{i.e.}, the accumulated bulk current through the crystals during the adiabatic evolving process from an initial state to a final state. In a quantum picture, the adiabatic current is essentially determined by the Berry phase of wave functions, \emph{i.e.},  the integral of the Berry curvature over the Brillouin zone \cite{xiao2010berry}. Therefore, for the adiabatic evolving process parameterized by a scalar $\lambda$, from the initial state ($\lambda = 0$) to the final state ($\lambda = 1$), the polarization difference $\Delta P_i$ can be written as \cite{resta1994macroscopic,PhysicsofFerroelectrics,xiao2010berry}
\begin{equation}
\Delta P_i = 2e\sum_n\int^1_0 d\lambda \int_{BZ} \frac{d\bm{k}}{(2\pi)^2} \Omega^n_{i,\lambda},
\end{equation}
where factor $2$ denotes the spin degrees of freedom, and the Berry curvature of valence-band wave function $u_n$ is defined by
\begin{equation}
\Omega^n_{i,\lambda} = i\left[\left\langle\frac{\partial u_n}{\partial k_i} \Bigg|\frac{\partial u_n}{\partial\lambda} \right \rangle-\left\langle\frac{\partial u_n}{\partial\lambda} \Bigg|\frac{\partial u_n}{\partial k_i} \right\rangle\right].
\end{equation}

Now let us first point out the mistakes made in Ref.~[\onlinecite{droth2016piezoelectricity}] and then present the correct results. To explore the piezoelectricity of \emph{h}-BN, Droth \emph{et al}. construct an adiabatic process parameterized by the band gap $\delta$  and then choose free graphene as the initial state with $\delta = 0$ (zero band gap) and \emph{h}-BN as the final state with $\delta = \Delta$ (where $\Delta$ is the band gap of h-BN). In this way, they write the polarization difference $\Delta P_i$ as \cite{droth2016piezoelectricity}
\begin{equation}
\Delta P_i = 2e\sum_\tau\int^\Delta_0 d\delta \int_{BZ/2} \frac{d\bm{q}}{(2\pi)^2} \Omega^\tau_{i,\delta},
\end{equation}
where $\tau$ denotes the $K$ $(\tau = 1)$ and $K^\prime$ $(\tau = -1)$ valleys, and the corresponding Berry curvature is \cite{droth2016piezoelectricity}
\begin{equation}
\Omega^\tau_{i,\delta} = i\left[\left\langle\frac{\partial u_\tau}{\partial q_i} \Bigg|\frac{\partial u_\tau}{\partial\delta} \right \rangle-\left\langle\frac{\partial u_\tau}{\partial\delta} \Bigg|\frac{\partial u_\tau}{\partial q_i} \right\rangle\right].
\end{equation}
Their chosen adiabatic process and parameter are incorrect for the piezoelectricity, because the correct adiabatic process of piezoelectricity should reflect the \emph{deformation-induced polarization difference} from the initial state of undeformed \emph{h}-BN to the final state of deformed \emph{h}-BN. In fact, such an adiabatic process for the piezoelectricity of any 2D piezoelectric crystal can be captured by the $\lambda$-dependent strain
\begin{equation}
\bm{\varepsilon}(\lambda) = \lambda\Delta\bm{\varepsilon} = \lambda{
\begin{pmatrix}
\Delta\varepsilon_{xx}&\Delta\varepsilon_{xy}\\
\Delta\varepsilon_{xy}&\Delta\varepsilon_{yy}
\end{pmatrix}}.
\end{equation}
Obviously, $\lambda = 0$ represents the initial state of the undeformed crystal, and $\lambda = 1$ denotes the final state of the deformed crystal. Consequently, the polarization difference along the $i$ direction is contributed by all components of the strain tensor and takes the form:
\begin{equation}
\Delta P_i = 2e\int^1_0 d\lambda\sum_{n,jk} \int_{BZ} \frac{d\bm{k}}{(2\pi)^2} \Omega^n_{i,jk}\Delta\varepsilon_{jk},
\label{eq:6}
\end{equation}
where the Berry curvature $\Omega^n_{i,jk}$ of the valance band with respect to $k_i$ and $\varepsilon_{jk}$ is
\begin{equation}
\Omega^n_{i,jk} = i\left[\left\langle\frac{\partial u_n}{\partial k_i} \Bigg|\frac{\partial u_n}{\partial\varepsilon_{jk}} \right \rangle-\left\langle\frac{\partial u_n}{\partial\varepsilon_{jk}} \Bigg|\frac{\partial u_n}{\partial k_i} \right\rangle\right].
\label{eq:7}
\end{equation}
Equation (\ref{eq:6}) can be further simplified as
\begin{equation}
\Delta P_i = \sum_{jk}\left[\sum_{n}2e\int_{BZ} \frac{d\bm{k}}{(2\pi)^2} \Omega^n_{i,jk}\right]\Delta\varepsilon_{jk}.
\end{equation}
The piezoelectricity is generally defined as a reversible linear electromechanical coupling effect, and the flexoelectricity is an electromechanical response of stress and strain gradients \cite{akinwande2017review}. Consequently, the piezoelectric coefficient, \emph{i.e.}, $e_{ijk} = \partial P_i/\partial \varepsilon_{jk}$, can be written as
\begin{equation}
e_{ijk} = 2e\sum_{n}\int_{BZ} \frac{d\bm{k}}{(2\pi)^2} \Omega^n_{i,jk} =  \frac{e}{\pi} \sum_{n} C^n_{i,jk}.
\label{eq:9}
\end{equation}
where $\Omega^n_{i,jk}$ is a short-hand notation of $\Omega^n_{i,jk}|_{\bm{\varepsilon}\rightarrow \bm{0}}$ (it will be used hereinafter), and $C^n_{i,jk}$ plays a similar role of the Chern number but is not quantized \cite{xiao2010berry}. Equation (\ref{eq:9}) is a general Berry phase expression of the piezoelectric coefficient for any 2D piezoelectric crystal and can be further extended to the case of three-dimensional crystals.

Next, we reformulate the analytical expression of $e_{222}$ for \emph{h}-BN. Because the strain-induced pseudomagnetic vector potential keeps the time-reversal invariant, the valley degenerate remains. Therefore, we can only consider the low-energy effective Hamiltonian of \emph{h}-BN at the $K$ valley in the continuum approximation, as follows:
\begin{equation}
H_K(\bm{q},\bm{\varepsilon}) = \Delta\sigma_z + \hbar v_F(\sigma_xq_x+\sigma_yq_y)+ \hbar v_F(\sigma_xA_x+\sigma_yA_y),
\label{eq:10}
\end{equation}
where $\hbar v_F = 3a\left|t\right|/2$, $\sigma_x$, $\sigma_y$ and $\sigma_z$ are the Pauli matrices, and the pseudomagnetic vector potential $\bm{A}$ as a perturbation in the Hamiltonian is given by \cite{neto2009electronic}
\begin{equation}
{\begin{pmatrix}
A_x \\
A_y
\end{pmatrix}} =
-\frac{3t\beta}{4\hbar v_F }{\begin{pmatrix}
\varepsilon_{xx}-\varepsilon_{yy} \\
-\varepsilon_{xy}
\end{pmatrix}},
\end{equation}
where $\varepsilon_{xy}= \partial_x u_y+\partial_y u_x$ is the shear strain. From the Hamiltonian in Eq.~(\ref{eq:10}), we can obtain the energy eigenvalues and normalized eigenstates for the conduction $(+)$ and valence $(-)$ bands of \emph{h}-BN without the strain perturbation, as follows:
\begin{equation}
\begin{gathered}
 E_\pm = \pm \sqrt{\hbar^2v_F^2(q_x^2+q_y^2)+\Delta^2},\\
 |u_\pm\rangle = \frac{1}{\sqrt{2E_+}}
 {\begin{pmatrix}
\frac{\hbar v_F (q_x-iq_y)}{\sqrt{E_+\mp\Delta}}\\
\pm\sqrt{E_+\mp\Delta}
\end{pmatrix}}.
\end{gathered}
\label{eq:12}
\end{equation}
Using the identity,
 \begin{equation}
 \left\langle\partial u_-/\partial R|u_+\right \rangle = \left\langle u_-|\partial H_K(\bm{q},\bm{\varepsilon}) / \partial R|u_+\right \rangle /(E_--E_+),
\end{equation}
with $R = q_i$ or $\varepsilon_{jk}$, one can rewrite the Berry curvature for the valence band of \emph{h}-BN in Eq.~(\ref{eq:7}) as
\begin{equation}
\Omega_{i,jk} = i\frac{\left\langle u_-|\partial H_K / \partial q_i|u_+\right\rangle \left\langle u_+|\partial H_K / \partial \varepsilon_{jk}|u_-\right\rangle-c.c.}{(E_+-E_-)^2},
\label{eq:14}
\end{equation}
where $c.c.$ denotes the complex conjugate. To obtain the Berry curvature $\Omega^K_{2,22}$ near the $K$ point, we first write the partial derivatives of Hamiltonian in Eq.~(\ref{eq:10}) with respect to $q_y$ and $\varepsilon_{yy}$ as
\begin{equation}
\partial H_K(\bm{q}, \bm{\varepsilon}) / \partial q_y = \hbar v_F \sigma_y, \partial H_K(\bm{q},\bm{\varepsilon}) / \partial \varepsilon_{yy} = \frac{3t\beta}{4} \sigma_x.
\label{eq:15}
\end{equation}
Substituting Eqs.~(\ref{eq:12}) and (\ref{eq:15}) into Eq.~(\ref{eq:14}), we can obtain the Berry curvature $\Omega^K_{2,22}$  near the $K$ point as follows:
\begin{equation}
\Omega^K_{2,22} = \frac{9at^2\beta\Delta}{16\left[ (3at/2)^2(q_x^2+q_y^2)+\Delta^2 \right]^{3/2}}.
\end{equation}
Because the piezoelectric coefficient $e_{222}$ of \emph{h}-BN is mainly contributed by both of the Berry curvatures in the vicinity of $K$ and $K^\prime$ points, the integral domain in Eq.~(\ref{eq:9}) can be replaced by two circles centered at $K$ and $K^\prime$ points, where the area of each circle should be equal to half of the first Brillouin zone in order to conserve the total number of states. Consequently, the piezoelectric coefficient $e_{222}$ reads
\begin{equation}
e_{222} = \frac{4e}{(2\pi)^2} \int_0^{2\pi} d\theta \int_0^{q_m} \frac{9at^2\beta\Delta qdq}{16\left[ (3at/2)^2q^2+\Delta^2 \right]^{3/2}},
\label{eq:17}
\end{equation}
where the factor $4$ contains both the spin and valley degenerates, $q = \sqrt{q_x^2+q_y^2}$, and $\pi q_m^2 = S_{BZ}/2$ with the area of the first Brillouin zone $S_{BZ} = 8\pi^2/{3\sqrt{3}a^2}$. We simplify Eq.~(\ref{eq:17}) and finally obtain the piezoelectric coefficient
\begin{equation}
e_{222} = \frac{e\beta}{2\pi a} \left( 1-\frac{\Delta}{\sqrt{\sqrt{3}\pi t^2+\Delta^2}}\right).
\label{eq:18}
\end{equation}
We adopt the same standard values of related parameters used in Ref.~[\onlinecite{droth2016piezoelectricity}] for \emph{h}-BN, \emph{i.e.}, $a = 1.44 $ \AA, $t=-2.3$ eV, $\Delta = 3.0$ eV, and $\beta = 3.3$. The calculated piezoelectric coefficient $e_{222}$ is $2.99\times 10^{-10}C/m$. If we further consider that the $\pi$-band electrons only contribute to 80\% of the piezoelectricity for \emph{h}-BN \cite{sai2003microscopic}, we will obtain the piezoelectric coefficient $e_{222}$ with its value $3.74\times10^{-10}C/m$, which agrees well with the result $(e_{222} = 3.71\times10^{-10}C/m)$ obtained from the first-principles calculations \cite{duerloo2012intrinsic}. Usually, the piezoelectric coefficient $e_{ijk}$ is defined by $e_{ijk} = \partial P_i/\partial \varepsilon_{jk}$ in Voigt notation, where $\varepsilon_{jk}$ is the strain tensor component with $\varepsilon_{xx}=u_{xx}$, $\varepsilon_{yy}=u_{yy}$, and $\varepsilon_{xy}=2u_{xy}$ for 2D system. For this definition of piezoelectric coefficient, the $\bar{6}m2$ point group in hexagonal crystal system requires
\begin{equation}
e_{211} = -e_{222} = e_{112} = e_{121}.
\label{eq:sym}
\end{equation}
However, if the piezoelectric coefficient $e_{ijk}$ is written as $e_{ijk} = \partial P_i/\partial u_{jk}$, where $u_{jk}=(\partial_i u_j+\partial_j u_i)/2$, with $\bm{u}$ as the in-plain displacement, Eq.~(\ref{eq:sym}) needs to be rewritten as
\begin{equation}
e_{211} = -e_{222} = \frac{1}{2} e_{112} = \frac{1}{2} e_{121}.
\end{equation}
Finally, we note that a recent analytical result of the piezoelectric coefficients for \emph{h}-BN is also obtained \cite{Rostami2017}. It should also be emphasized that Eq.~(\ref{eq:18}) in our comment is only for \emph{h}-BN.

In conclusion, we formulate a general expression for the piezoelectric coefficients of 2D piezoelectric crystals by means of the Berry phase approach. Using the effective Hamiltonian in the continuum approximation, we analytically derive the expression for the piezoelectric constants of \emph{h}-BN. The analytical result is in complete consistency with the first-principles calculations.

This work was supported financially by National Natural Science Foundation of China under Grant Nos 11502308, 11472313 and 11572355, Guangdong Natural Science Foundation of China under Grant No. 2016A030310205, and the fundamental research funds for the central universities under Grant No. 17lgpy31.


\end{document}